\documentclass[12pt]{iopart}

\usepackage{iopams}

\newcommand{\pfaff}{\mbox{ Pfaff }}
\begin{document}

\title[Symplectic Structure of the Real Ginibre Ensemble]{Symplectic Structure of the Real Ginibre Ensemble}

\author{Hans-J\"{u}rgen Sommers}

\address{\it Fachbereich Physik, Universit\"{a}t Duisburg-Essen \\
47048 Duisburg, Germany
} \ead{H.J.Sommers@uni-due.de}
\begin{abstract}
We give a simple derivation of all $n$-point densities for the eigenvalues of the real Ginibre ensemble with even dimension $N$ as quaternion determinants. A very simple symplectic kernel governs both, the real and complex correlations. 1-and-2-point correlations are discussed in more detail. Scaling forms for large dimension $N$ are derived.
\end{abstract}

 \submitto{\JPA}
 
\section{Introduction}

More than 40 years ago Ginibre \cite{Ginibre65} proposed three types of general Gaussian non-Hermitian matrix ensembles, those with complex, quaternion real and real entries. While he already gave the joint probability density (jpd) of eigenvalues and correlations for the first two cases, the last case was harder to solve. It started  with the presentation of the  jpd by Lehmann and the author 25 years later \cite{Lehmann91}. Edelman rederived the jpd and determind the density of complex eigenvalues in 1997 \cite{Edelman97}. In more recent publications Kanzieper and Akemann presented the complex correlations as Pfaffians \cite{Kanzieper05}, Sinclair derived a generating Pfaffian functional \cite{Sinclair06} and finally Forrester and Nagao were able to determine the real and complex correlations as Pfaffians with the help of orthogonal polynomials \cite{Forrester07}. The last step should be considered as the solution of the problem of correlations of the real Ginibre ensemble.  Here we give an alternative simple derivation and the generalisation to arbitrary real, complex or crossed correlations. \\
In this paper we present the general $n$-point densities, all entries real or complex, as quaternion determinant governed by the simple kernel
\begin{equation}
\label{-1}
K_N (z,z') = \frac{z - z'}{2 \sqrt{2 \pi}} \sum_{n=0}^{N-2} \frac{(z z')^n}{n!} \; .
\end{equation}
We restrict ourselves to even dimension $N$. This is the only place where the matrix dimension $N$ occurs. The power expansion of (\ref{-1}) generates the inverse of an $N$-dimensional submatrix ($1\leqslant k, l\leqslant N$) of the infinite dimensional matrix
\begin{equation}
A_{kl} = \int d^2z_1 \int d^2z_2 \; {\cal F} (z_1,z_2) z_1^{k-1} z_2^{l-1}
\end{equation}
where ${\cal F} (z_1,z_2)$ is a skew-symmetric complex measure reflecting the symplectic structure of the real Ginibre ensemble. A simple argument without use of orthogonal polynomials yields directly equation (\ref{-1}). Furthermore we discuss properties of 1-and-2-point functions in more detail.\\
The real asymmetric Gaussian ensemble has many applications in physics and social sciences as biological webs \cite{May72}, neural networks \cite{Sompolinsky88}, directed quantum chaos \cite{Efetov97}, financial markets \cite{KWAPIEN06}. The probability density of matrices $J_{ij}$ ($1\leqslant i, j\leqslant N$) we are talking about in this paper is simply $\propto \exp \bigl(-\sum_{ij} J_{ij}^2/2\bigr)$. And the eigenvalues $z_i$ of the matrix $J_{ij}$ are the zeros of the characteristic polynomial $\det (J_{ij} - z) = 0$ and therefore real or pairwise complex conjugate.
\section{Symplectic Structure}
We start with the joint probability density of eigenvalues for the real Ginibre ensemble derived by Lehmann and Sommers \cite{Lehmann91} and calculate from it a generating functional for correlation functions \cite{Sinclair06,Forrester07}:
\begin{equation}
\label{1}
Z[f] = \int d^2z_1 \ldots d^2z_N \; P(z_1,z_2,\ldots,z_N) f(z_1) \ldots f(z_N)\; .
\end{equation}
$P$ contains $\left[ \frac{N}{2} \right] +1$ pieces with $R$ real eigenvalues ($0\leqslant R\leqslant N$) and $Q = (N-R)/2$ complex conjugate pairs in a chosen order. We extend here for convenience the eigenvalues to complex values, $z_k = x_k + i y_k$. Thus for real eigenvalues, $z = x_k$, $P(\ldots)$ is concentrated on the real axis: $y_k = 0$. Afterwards we may symmetrize $ P(z_1,z_2,\ldots,z_N)$ with respect to all permutations of $z_k$ and integrate all variables $z_k$ over the whole complex plane dividing the result by $N!$. $P(\ldots)$ is proportional to the Vandermonde determinant
\begin{equation}
\prod_{i>j} (z_i - z_j) = \det \left(z_1^{k-1},\ldots, z_N^{k-1}\right)
\end{equation}
which has a definite sign for the chosen order. Using the method of alternating variables \cite{Mehta04} we may then integrate out (\ref{1}) and obtain a Pfaffian. The simplest way to derive it, is using Grassmannians, see the discussion at the end of this paper. For even $N$ the result is 
\begin{equation}
\label{3}
Z[f] = {\cal N} \pfaff (\widetilde{A}_{kl}) = {\cal N} \sqrt{\det (\widetilde{A}_{kl})}
\end{equation}
with a skew-symmetric matrix
\begin{equation}
\label{4}
\widetilde{A}_{kl} = \int d^2 z_1 d^2 z_2 f(z_1) f(z_2) {\cal F} (z_1, z_2) z_1^{k-1} z_2^{l-1} 
\end{equation}
where ${\cal F} (z_1, z_2)$ is a skew-symmetric function $({\cal F} (z_1, z_2) = - {\cal F} (z_2, z_1))$ of two complex variables $z_1,z_2$
\begin{eqnarray}
\label{5}
\nonumber
\fl {\cal F} (z_1, z_2)  = {\rm e}^{-(z_1^2 + z_2^2)/2} \left[ 2 i \delta^2 (z_1 - \bar{z}_2) \{ \Theta(y_1) \mbox{ erfc} (y_1\sqrt{2}) \right. \\
 - \left. \Theta(y_2) \mbox{ erfc} (y_2 \sqrt{2}) \} + \delta(y_1) \delta(y_2) \bigl(\Theta (x_2 - x_1) - \Theta (x_1 - x_2) \bigr) \right]\; .
\end{eqnarray}
${\cal N}$ is a normalization constant, which can be restored afterwards. It is related to the volume of the real orthogonal group \cite{Lehmann91}.
Actually the variables $z_1, z_2$ are considered here as two dimensional real vectors. This symplectic measure reveals two contributions, one from the real axis with $y_1 = y_2 = 0$ and one from the complex plane off the real axis with $\delta^2(z_1- \bar{z}_2) = \delta (x_1 - x_2) \delta(y_1 + y_2)$. The step functions $\Theta(\ldots)$ reflect the chosen order of eigenvalues. Functional derivatives of $Z[f]$ with respect to $f(z)$ at $f \equiv 1$ give immediately all $n$-point densities:
\begin{equation}
R_1(z_1) = \left. \frac{\delta Z[f]}{\delta f(z_1)} \right|_{f \equiv 1} \; , \quad
R_2(z_1,z_2) = \left. \frac{\delta^2 Z[f]}{\delta f(z_1) \delta f(z_2)} \right|_{f \equiv 1}\; ,\; \ldots
\end{equation}
Equations (\ref{1}) - (\ref{5}) are completely equivalent to the joined density of eigenvalues obtained by Lehmann and Sommers \cite{Lehmann91} and rederived by Edelman \cite{Edelman97}. We will use them to calculate the correlation functions (or $n$-point densities, which should be distinguished from connected correlations).\\
From (\ref{3}) we obtain
\begin{equation}
\label{7}
\frac{\delta Z[f]}{\delta f (z_1)} = Z[f] \int d^2z_2 \; f(z_2) {\cal F} (z_1, z_2) \widetilde{K} (z_2, z_1)
\end{equation}
with the kernel
\begin{equation}
\widetilde{K} (z_2, z_1) = \sum_{k,l}^{1\ldots N} \widetilde{A}_{kl}^{-1} z_2^{k-1} z_1^{l-1} \; . 
\end{equation}
At $f \equiv 1$ we have $\widetilde{A}_{kl} = A_{kl}$, $\widetilde{K} (z_2, z_1) = K (z_2,z_1)$ and $Z[1] = 1$. Thus
\begin{equation}
\label{9}
R_1(z_1) = \int d^2 z_2 {\cal F} (z_1, z_2) K(z_2, z_1)\; .
\end{equation}
$R_1(z_1)$ contains two independent parts, one on the real axis and one in the complex plane. Comparing the latter with Edelman's expression for the density in the complex plane (not on the real axis) \cite{Edelman97} we obtain (since $z_k$ and the complex conjugate $\bar{z}_{k}$ can be considered as independent variables)
\begin{equation}
\label{10}
K(z_2,z_1) = \frac{z_2 - z_1}{2 \sqrt{2 \pi}} \sum_{n=0}^{N-2} \frac{(z_1 z_2)^n}{n!} = \sum_{k,l}^{1 \ldots N} A_{kl}^{-1} z_2^{k-1} z_1^{l-1} \; .
\end{equation}
Thus, remarkably, the inverse $N$-dimensional skew-symmetric matrix $A_{kl}^{-1}$ has a very simple tridiagonal form and the bulk part and the real-axis part of the complex density $R_1(z)$ are intimately related via the kernel $K(z_2,z_1)$. Equation (\ref{10}) can independently be checked by calculating $A_{kl}^{-1}$ directly from (\ref{4}) and (\ref{5}). Now, not only we recover Edelman's bulk part of $R_1(z)$ \cite{Edelman97}, but also we obtain the density on the real axis  first obtained by Edelman, Kostlan and   
 Shub\cite{Edelman94}
\begin{equation}
R_1(z) = R_1^C (z) + \delta(y) R_1^{R}(x)
\end{equation}
with
\begin{equation}
R_1^C(z) = {\rm e}^{-(x^2 - y^2)} \frac{2 \sqrt{2}}{\pi} \int\limits_{|y| \sqrt{2}}^{\infty} du \; {\rm e}^{-u^2} |y| \sum_{n=0}^{N-2} \frac{|z|^{2n}}{n!}
\end{equation}
and
\begin{equation}
\label{11}
R_1^R (x) = \int\limits_{-\infty}^{+ \infty} \frac{dx' \; | x - x' |}{2 \sqrt{2 \pi}} {\rm e}^{-(x^2 + x'^2)/2} \sum_{n=0}^{N-2} \frac{(x x')^n}{n!}
\end{equation}
Writing 
\begin{equation}
\label{12}
{\rm e}^{- x x'} \sum_{n=0}^{N-2} \frac{(x x')^n}{n!} = \int\limits_{x x'}^{\infty} du \; {\rm e}^{-u} \frac{u^{N-2}}{(N-2)!}
\end{equation}
we see that expression (\ref{11}) is positive for even $N\geqslant 2$, for which it was derived. For odd $N$ there is a correction term, which ensures positivity. Equation (\ref{12}) makes it easy to derive large $N$-expansions by saddle-point integration. The simplest results for large $N$ are the Girko circle law \cite{Girko84,Sommers88}
\begin{equation}
\label{13}
R_1^C (z) \simeq \frac{1}{\pi} \Theta (\sqrt{N} - |z|)
\end{equation}
and a constant density on the real axis \cite{Forrester07}
\begin{equation}
\label{14}
R_1^R (x) \simeq \frac{1}{\sqrt{2\pi}} \Theta (\sqrt{N} - |x|) \; .
\end{equation}
This is in contrast to the semicircle density for a Gaussian Hermitian matrix (Wigner semicircle law). A new type of behaviour is expected near the edge. The average number of real eigenvalues $N_R (N)$ is in agreement with Edelman \cite{Edelman97}:
\begin{equation}
N_R (N) = 1 + \frac{\sqrt{2}}{\pi} \int\limits_{0}^{1} \frac{dt \; t^{1/2} ( 1 - t^{N-1})}{(1-t)^{3/2} (1+t)} \simeq \sqrt{\frac{2N}{\pi}}
\end{equation}
For calculating the next correlation functions one has to take functional derivatives of (\ref{7}), for which we use
\begin{eqnarray}
\label{16}
\nonumber
 \delta \widetilde{K} (z_2,z_1) =& \int d^2z_3 \delta f(z_3) \int d^2 z_4 f(z_4) {\cal F} (z_3, z_4) \cdot \\
&\left[ \widetilde{K}(z_2, z_4) \widetilde{K}(z_3, z_1) - \widetilde{K}(z_2, z_3) \widetilde{K}(z_4, z_1) \right] \; .
\end{eqnarray}
This equation is enough to generate all higher functional derivatives. Equation (\ref{16}) implies for the 2-point density

\begin{eqnarray}
\label{17}
\nonumber
\fl R_2(z_1,z_2) = {\cal F} (z_1,z_2)  K(z_2,z_1) + \int d^2 z_3 \int d^2 z_4 {\cal F} (z_1,z_3) {\cal F} (z_2,z_4) \cdot\\
\left\{ K (z_3, z_1) K(z_4, z_2) + K(z_3, z_4) K(z_2, z_1) - K(z_3, z_2) K(z_4, z_1) \right\} \; .
\end{eqnarray}
At the end of this paper we will write this as a quaternion determinant.
\section{2-point densities}
Let us single out in $R_2(z_1,z_2)$ the nonsingular bulk part $R^C_2(z_1,z_2)$, that contains no $\delta$-contributions and describes correlations in the complex plane off the real axis \cite{Kanzieper05}
\begin{eqnarray}
\label{18}
\nonumber
 R_2^C (z_1, z_2) =& 4 \mbox{ sgn}(y_1) \mbox{ erfc} (|y_1| \sqrt{2}) \mbox{ sgn}(y_2) \mbox{ erfc} (|y_2| \sqrt{2}) \cdot\\
 \nonumber
& \cdot {\rm e}^{-(z_1^2 + \bar{z}_1^2 + z_2^2 + \bar{z}_2^2)/2} \left\{ - K(\bar{z}_1,z_1) K(\bar{z_2},z_2) - \right. \\
& \left. - K(\bar{z}_1,\bar{z}_2) K(z_2,z_1) + K(\bar{z}_1,z_2) K(\bar{z_2},z_1) \right\}
\end{eqnarray}
For small $z_1, z_2$ this behaves as $16 |y_1 y_2| (x_1^2 + x_2^2) > 0$. For $(z_1, \bar{z}_1) \rightarrow (z_2, \bar{z}_2)$ the correlation  $R_2^C (z_1,z_2)$ behaves $\propto |z_1 - z_2|^2$. On the other hand, the most singular part of $R_2(z_1,z_2)$ on the real axis is $\delta (y_1) \delta (y_2) R_2^R (x_1,x_2)$ where $R_2^R (x_1,x_2)$ describes correlations of real eigenvalues and is given by:
\begin{eqnarray}
\label{19}
\nonumber
\fl R_2^R (x_1,x_2) = {\rm e}^{-(x_1^2 + x_2^2)/2} \left\{ \mbox{ sgn} (x_2 - x_1) K(x_2,x_1) + \right.\\
\nonumber
\int dx_3 \int dx_4 \; {\rm e}^{-(x_3^2 + x_4^2)/2} \mbox{ sgn}(x_3 - x_1) \mbox{ sgn} (x_4 - x_2) \cdot\\
\cdot \left. \left[ K(x_3,x_1) K(x_4,x_2) + K(x_3,x_4) K(x_2,x_1) - K(x_3,x_2) K(x_4,x_1) \right] \right\} \; .
\end{eqnarray}
Both expressions (\ref{18}) and (\ref{19}) are contained in results of \cite{Forrester07}. 
We proceed now to derive scaling forms of the correlations in the large-$N$ limit. Since the densites (\ref{13}) and (\ref{14}) in the complex plane and on the real axis are constant, the scaled correlation functions are simply obtained by taking the $N \rightarrow \infty$ limit of equation (\ref{10})
\begin{equation}
K(z_2, z_1) \simeq \frac{z_2 - z_1}{2 \sqrt{2 \pi}} {\rm e}^{z_1 z_2} \; .
\end{equation}
Obviously the  symplectic measure function ${\cal F}(z_1,z_2)$ (\ref{5}) does not depend on $N$. This implies for the 1-point densities
\begin{equation}
R_1^C (z) \simeq \sqrt{\frac{2}{\pi}} {\rm e}^{2 y^2} |y| \mbox{ erfc}(|y| \sqrt{2})
\end{equation}
and
\begin{equation}
R_1^R(x) \simeq \frac{1}{\sqrt{2 \pi}} \; .
\end{equation}
Both are independent of $x$ and only for $|y| \gg 1$, i.e. far from the real axis, $R_1^C (z) $ goes to (\ref{13}) . This is generally valid: far from the real axis the correlations go to Ginibre's (i.e. the Ginibre ensemble for general complex matrices \cite{Ginibre65}). For example the bulk part $R_2^C (z_1,z_2)$ of the 2-point correlation function for large $N$ is given by 
\begin{eqnarray}
\label{23}
\nonumber
R_2^C (z_1,z_2) \simeq& R_1^C(z_1) R_1^C (z_2) \frac{1}{4 y_1 y_2} \Bigl\{ 4 y_1 y_2 \Bigl. \\
\nonumber
&+ \bigl( (x_1 - x_2)^2  + (y_1 - y_2)^2\bigr) {\rm e}^{-(x_1 - x_2)^2 - (y_1 + y_2)^2}\\
&\left. - \bigl( (x_1 - x_2)^2  + (y_1 + y_2)^2\bigr) {\rm e}^{-(x_1 - x_2)^2 - (y_1 - y_2)^2} \right\}
\end{eqnarray}
Far from the real axis only  the Ginibre result
\begin{equation}
R_2^C (z_1,z_2) \simeq \frac{1}{\pi^2} \left( 1 - {\rm e}^{-|z_1 - z_2|^2} \right)
\end{equation}
survives. Finally we compute the asymptotic form of the real 2-point correlations
\begin{equation}
\fl R_2^R (x_1, x_2) \simeq \frac{1}{2\pi} \left( 1 - {\rm e}^{-(x_1 - x_2)^2} \right) + \frac{|x_1 - x_2|}{2 \sqrt{2 \pi}} {\rm e}^{-\frac{(x_1 - x_2)^2}{2}} \mbox{ erfc} \left(\frac{|x_1 - x_2|}{\sqrt{2}} \right) \; .
\end{equation}
This is different from Wigner-Dyson correlations, but shows $\beta = 1$ level repulsion. For small distances we have $R_2^R (x_1, x_2) \simeq \frac{|x_1 - x_2|}{2 \sqrt{2 \pi}}$, for large distances we have $R_2^R (x_1, x_2)  \simeq \frac{1}{2\pi} \left( 1 - \frac{{\rm e}^{-(x_1 - x_2)^2}}{(x_1 - x_2)^2} \right)$, a much faster decay of connected  correlations than for GOE (the Gaussian orthogonal ensemble), more similar to those for the Ginibre ensemble in the complex plane. 
\section{$n$-point densities}
Finally we observe that the correlation (\ref{16}) can be written as a quaternion determinant of a selfdual matrix (related to a Pfaffian of a skew-symmetric matrix) and can be generalized for $n$-point densities ($k,l = 1, 2, \ldots, n$)
\begin{equation}
\label{26}
\fl R_n(z_1, \ldots, z_n) = \mbox{Qdet } \left[ \begin{array}{cc}
\int d^2z \; {\cal F} (z_k, z) K (z,z_l)& K(z_k,z_l)\\
-\widetilde{{\cal F}} (z_k, z_l)& \int d^2z \; K (z_k,z) {\cal F} (z, z_l) 
\end{array} \right]
\end{equation}
with
\begin{equation}
\widetilde{{\cal F}} (z_1, z_2) = {\cal F} (z_1, z_2) - \int d^2z d^2z' {\cal F} (z_1,z) K(z,z') {\cal F} (z', z_2) \; .
\end{equation}
The sign of the quaternion determinant is given by the diagonal elements\\ $R_1(z_1), \ldots , R_n(z_n)$ (see (\ref{9})). The term $R_1(z_1)\cdot R_2(z_2) \cdots R_n(z_n)$ appearing in the expansion of the quaternion determinant just gives the asymptotics of $R_n (z_1, \ldots, z_n)$ for large separation.\\
For an alternative derivation one may write the generating functional $Z[1 + u]$ as a Gaussian integral over Grassmannians $\zeta$
\begin{equation}
\fl Z[1 + u] = \int D\zeta \exp\left( - \frac{1}{2} \int d^2z \int d^2z' \xi(z) \xi(z') 
{\cal F} (z,z') (1 + u(z)) (1+ u(z'))\right)
\end{equation}
with Grassmannian fields  $\xi(z)$ with $\langle \xi(z) \xi (z')\rangle_{u=0} = K(z,z')$ and expand $Z[1+u]$ using the fermionic Wick theorem to obtain all correlation functions. The entries in (\ref{26}) are the building elements of the diagrammatic expansion. A similar expansion for the quaternion determinants proves the claim (\ref{26}). One crucial point is that although the quaternion determinant is a square root, it is an analytic function of the entries.
\section{Conclusion}
In conclusion, without use of orthogonal polynomials we have given a transparent derivation of the correlations for the real Ginibre ensemble, which are governed by one simple symplectic kernel $K(z,z')$ describing a deep connection between real and complex correlations. Real eigenvalues, which meat at a point on the real axis, continue to move along the imaginary direction repelling each other without changing the real coefficients of their characteristic polynomial. Calculations have been done for even $N$. The algebra for odd $N$ is more complicated. However one may conjecture, that the $n$-point densities for odd $N$ can be obtained by analytic continuation from even $N$ preserving positivity. Moreover we have discussed some properties of 1-and-2-point functions in detail. A lot of further properties can be discussed in future.\\
After completing most of this paper, I have seen some recent papers \cite{Kanzieper05,Sinclair06,Forrester07}, where many of the results reported here are partly contained. However this paper contains in addition some generalisations and tries to make transparent the beauty of the structure of the real Ginibre ensemble, which also may be useful in many fields of sciences.\\
The author acknowledges support by SFB/TR12 of the Deutsche Forschungsgemeinschaft.


\section*{References}
\begin{thebibliography}{10}
\bibitem{Ginibre65} J.Ginibre, J. Math. Phys. {\bf 6}, 440 (1965)
\bibitem{Lehmann91} N.Lehmann and H.-J. Sommers, Phys. Rev. Lett. {\bf 67}, 941 (1991)
\bibitem{Edelman97} A.Edelman, J. Multivariate Analisis {\bf 60}, 203 (1997)
\bibitem{Kanzieper05} E.Kanzieper and G. Akemann, Phys. Rev. Lett. {\bf 95}, 230201 (2005) 
\bibitem{Sinclair06} C.D.Sinclair, arXiv: math-ph/0605006 (2006)
\bibitem{Forrester07} P.J.Forrester and T. Nagao, to be published (2007)
\bibitem{May72} R.M.May, Nature {\bf 298 } 413 (1972)
\bibitem{Sompolinsky88} H.Sompolinsky, A.Crisanti and H.-J.Sommers, Phys. Rev. Lett. {\bf 61} 259 (1988) 
\bibitem{Efetov97} K.B.Efetov, Phys. Rev. Lett. {\bf 79} 491 (1997)
\bibitem{KWAPIEN06} J.Kwapien, S. Drozdz, A.Z. Gorski and  F. Oswiecimka, Acta  Phys. Pol. {\bf B37}, 3039 (2006)
\bibitem{Edelman94} A. Edelman, E. Kostlan and M. Shub, J. Amer. Math. Soc. {\bf7}, 247 (1994) 
\bibitem{Mehta04} M.L.Mehta, Random Matrices (Amsterdam Elsevier) 2004
\bibitem{Girko84} V.L.Girko, Theory Probab. Appl. {\bf 29} 694 (1984)
\bibitem{Sommers88} H.-J.Sommers, A.Crisanti, H.Sompolinsky and Y.Stein, Phys. Rev.Lett {\bf 60} 1895 (1988)
\end{thebibliography}
\end{document}